\documentclass[amsmath,amssymb,prx,twocolumn,superscriptaddress,tightenlines,twocolumn,showpacs,floatfix,longbibliography,nofootinbib]{revtex4-2}
\pdfoutput=1
\usepackage{graphicx}
\usepackage{dcolumn}
\usepackage{bm}
\usepackage{color}

\usepackage{amsmath}
\DeclareMathOperator{\sign}{sign}

\definecolor{nblue}{RGB}{28,130,185}
\definecolor{cgreen}{RGB}{76,153,0}
\definecolor{myorange}{RGB}{245,156,74}

\usepackage{hyperref}
\hypersetup{
  colorlinks=true,
  citecolor=cyan,
  urlcolor=cgreen
}

\def \beq{\begin{eqnarray}}
\def \eeq{\end{eqnarray}}

\newcommand{\bk}{\bm{k}}

\newcommand{\br}{\bm{r}}
\newcommand{\bq}{\bm{q}}
\newcommand{\iu}{{i\mkern1mu}}

\begin{document}

\title{Effervescent waves in a binary mixture with non-reciprocal couplings}

\author{Suropriya Saha}
\affiliation{Max Planck Institute for Dynamics and Self-Organization (MPIDS), D-37077 G\"ottingen, Germany}

\author{Ramin Golestanian}
\email{ramin.golestanian@ds.mpg.de}
\affiliation{Max Planck Institute for Dynamics and Self-Organization (MPIDS), D-37077 G\"ottingen, Germany}
\affiliation{Rudolf Peierls Centre for Theoretical Physics, University of Oxford, Oxford OX1 3PU, United Kingdom}

\date{\today}

\begin{abstract}
Non-reciprocal interactions between scalar fields that represent the concentrations of two active species are known to break the parity and time-reversal (PT) symmetries of the equilibrium state, as manifested in the emergence of travelling waves. We explore the notion of nonlinear non-reciprocity and consider a model in which the non-reciprocal interactions can depend on the local values of the scalar fields. For generic cases where such couplings exist, we observe the emergence of spatiotemporal chaos in the steady-state. We associate this chaotic behaviour with a local restoration of PT symmetry in fluctuating spatial domains, which leads to the coexistence of oscillating densities and phase-separated droplets that are spontaneously created and annihilated. We uncover that this phenomenon, which we denote as {\em effervescence}, can exist as a dynamical steady-state in large parts of the parameter space in two different incarnations, as characterized by the presence or absence of an accompanying travelling wave. 
\end{abstract}
\maketitle

\section{Introduction}\label{sec:intro}

Interactions between components of biological and artificial living matter are mediated in a wide variety of ways across the scales \cite{Gompper2020}: from complex behaviour patterns in humans \cite{Rio2018}, to visual perception in birds \cite{Ballerini2008}, hydrodynamic interactions in ensembles of cilia and flagella \cite{Uchida2010_PRL,Brumley2014}, information-controlled feedback in programmable active colloids \cite{LavergneBechinger2019, Khadka2018}, and chemical fields in catalytically active colloids \cite{SotoRaminPRL14,Soto2015,sahaNJP_2019} and enzymes \cite{agudo2018enhanced,Jaime2019}. These microscopic interactions quite generically break action-reaction symmetry due to non-equilibrium conditions. Reciprocity breaking has already had a far reaching impact in fields like structural mechanics, in realizing meta-materials \cite{Coulais2017}, and in optics, by achieving photon blockade \cite{PhotonBlocadePRL}. In recent years, non-reciprocity in interactions has generated interest as an exciting new ingredient to develop minimal models for active matter systems out of thermodynamic equilibrium \cite{Jaime2019,sahaPRX_2020,You_Marchetti_2020,Klapp_Loos_2020,Fruchart2021,ouazan-reboul2021Nonequilibrium, LokrshiMaitraPRE}. 

\begin{figure*}[t]
	\centering
	\includegraphics[width= 0.85\linewidth]{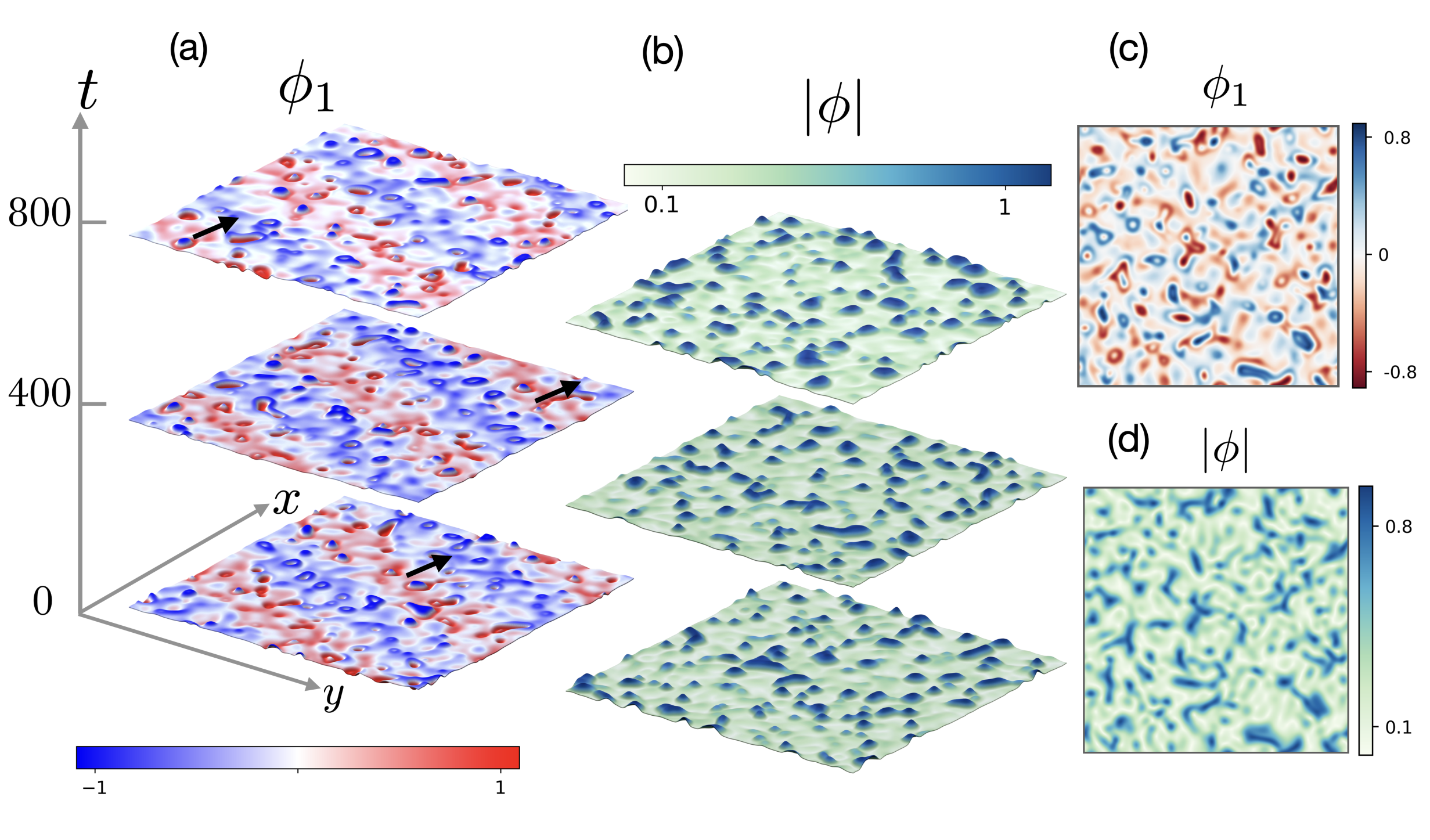}
	\caption{Emergence of effervescence in the dynamical steady-states of the nonlinear NRCH model. 
	Panel (a) shows snapshots of the field $\phi_1$ at the times indicated, which demonstrate the effervescent waves, corresponding to $\alpha_0=4$ and $\alpha_1=5$ (see Supplemental Movie S1). The black arrows mark the progression of the travelling wave. Panel (b) shows the corresponding amplitude field $|\phi|$. Note that while (a) shows a clear striped pattern, $|\phi|$ deviates from a constant value in areas where fluctuating droplets are formed. For $\alpha_1$ sufficiently larger than $\alpha_0$, we observe an effervescent steady-state without an accompanying travelling wave. Panels (c-d) shows snapshots of $\phi_1$ and $|\phi|$ from simulations corresponding to $\alpha_0 = 2.3$ and $\alpha_1 = 4.6$ (see Movie S2 for the full time evolution).} 
	\label{fig:nrch_SS}
\end{figure*}


Conserved active scalar field theories for two species with non-reciprocal interactions display travelling waves, moving patterns and oscillations in the steady-state \cite{sahaPRX_2020, You_Marchetti_2020}. When activity, i.e. the strength of non-reciprocity, is strong enough to win over the thermodynamic forces driving the system towards bulk phase separation, the system reaches novel steady-states that break the parity and time-reversal symmetry of the bulk-separated equilibrium state. The transition to travelling patterns occurs upon tuning the parameters of the model such that an exceptional point is crossed \cite{Heiss_2012,sahaPRX_2020,Fruchart2021}, at which the eigenvalues of the dynamical matrix determining the stability of the fully mixed state to small perturbations acquire complex values. Moreover, the pair of eigenvalues coalesce at the exceptional point and the corresponding eigenvectors become parallel \cite{Heiss_2012}. This transition from a state with a spontaneously broken symmetry to a state where the broken symmetry is restored also occurs in non-reciprocally coupled polar flocks with a non-conserved vector order parameter and in phase-synchronization models \cite{LokrshiMaitraPRE, Fruchart2021}. 
Similar phenomena are frequently encountered in open quantum systems that can dissipate or absorb energy by interacting with their surroundings so that they are described by a non-hermitian Hamiltonian \cite{Maghrebi-prx-2020}. So far, photonic systems have been primarily used for experimental realizations of non-hermitian quantum mechanics \cite{Longhi_PRL.105.013903, Liu2021}. However, classical systems have the potential to be used for similar purposes, for example using coupled enzyme cycles in reaction networks \cite{Tang2021}. 


Here we have generalized the non-reciprocal Cahn-Hilliard (NRCH) model developed in \cite{sahaPRX_2020} for two non-reciprocally interacting conserved species with the aim to explore the pattern forming ability of the scalar field theories with non-reciprocal interactions beyond travelling waves. While inspiration is abundant in the rich literature on pattern formation \cite{kuramoto2003chemical, CrossHohenberg_RMP.65.851} and chemical turbulence \cite{Aranson_RevModPhys.74.99, Chate1996}, in the past decade or so it has been established in the field of active matter that non-equilibrium local activity can drive turbulent flows. Examples of such phenomena include the theoretical prediction that ciliary carpets show phase ordering or proliferating turbulent spiral waves due to long-ranged hydrodynamic interactions \cite{Uchida2010_PRL}. Turbulence has been established in experimental systems consisting of concentrated bacterial suspensions \cite{Dunkel_Goldstein_PRL228102}, 
active nematic systems with defect-unbinding dynamics \cite{Sanchez2012,Berta-prx,sumesh-prl-2013}, and protein propagation patterns on curved membranes \cite{Tan2020}. While many of the examples listed here concern systems without explicit number conservation, recent developments in active phase separation have highlighted the significance of number conservation in active matter systems \cite{WittkowskiNJP2017, GenThermoSolon18, ClusterBubblyTjhung18,Frey-prx-2020}.  

Using theoretical analysis and simulations, we show that an interplay between linear and nonlinear non-reciprocal interactions produces travelling waves, spatiotemporal chaos without any apparent structure and a hybrid state in which {\it effervescent waves}, namely, waves with phase-separated droplets of matter in their wake, are generated (see Fig. \ref{fig:nrch_SS}, videos S1 and S2 in the supplement).
The outline of the paper is as follows. First, we will introduce the phenomenon of effervescence and summarize our findings in Sec. \ref{sec:effer}. We then introduce our theoretical model in Sec. \ref{sec:model} explaining the choice of a particular free energy that is invariant under unitary rotations in the composition plane, which makes the equations tractable to theoretical analysis. In Sec. \ref{sec:wave} we study the plane wave solutions that enable us to determine, via a stability analysis, the region of the parameter space in which these waves are unstable. The unstable region of the parameter space is then probed in Sec. \ref{sec:state-diag} by solving the equations of motion numerically to classify the dynamical steady-states and summarize the results in a state diagram. Finally, the effect of composition is studied in Sec. \ref{sec:compos}, which is followed by concluding remarks in Sec. \ref{sec:conc}. Some details of the linear stability analysis are relegated to Appendix \ref{appA:stability}.

\section{Effervescence}\label{sec:effer}

\begin{figure}[t]
	\centering 
	\includegraphics[width= 1. \linewidth]{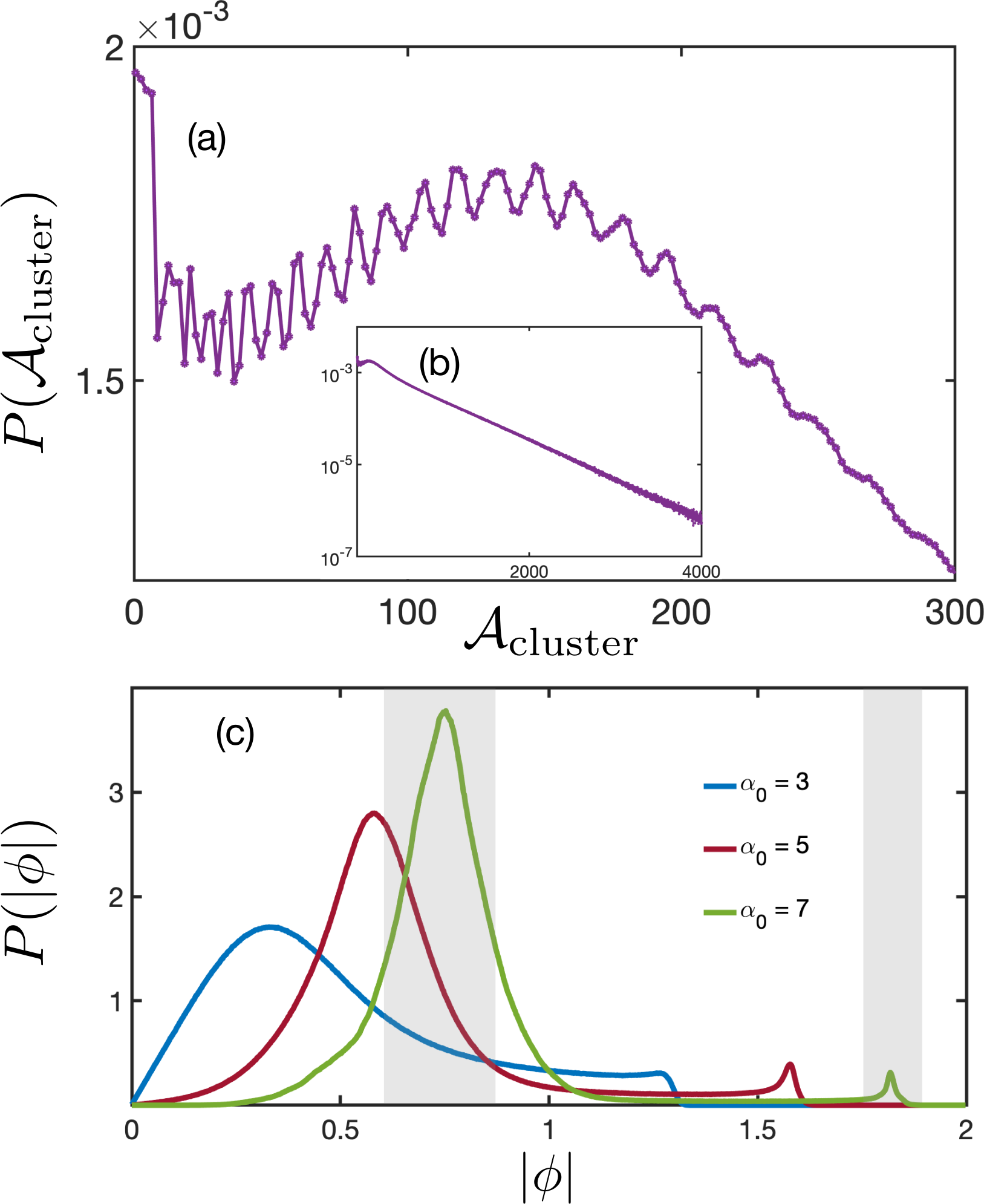}
	\caption{Effervescence with size-selected reciprocal granules. (a) Probability distribution for the area of the clusters made of reciprocal granules ${\cal A}_{\rm cluster}$ (in units of $q_0^{-2}$). The peak at the origin reveals the existence of a large population of droplets, and the oscillatory motif of the full distribution shows that the clusters are formed as aggregates of a number of similarly sized granules. (b) The observed exponential decay of the cluster size distribution (as evident in the semi-log scale) verifies the existence of an underlying size selection mechanism. (c) Probability distribution for the modulus of the parameter field $|\phi|$ in steady state, revealing the coexistence of traveling waves with condensates in which the effective non-reciprocal interaction vanishes, i.e. the reciprocal granules. }
	\label{fig:granule-size}
\end{figure}


An emergent feature of the non-reciprocal interactions implemented at the linear level in a mixture of two species is the spontaneous breaking of space-translation, time-translation, time-reversal, and polar symmetries, through the formation of traveling patterns \cite{sahaPRX_2020,You_Marchetti_2020}. When introducing nonlinear terms in the non-reciprocal interaction, in the spirit of a Landau expansion, we observe spontaneous creation and annihilation of droplets in combination with the traveling pattern, or even, in its absence, as shown in Fig. \ref{fig:nrch_SS}. We find droplets enhanced in either species 1 or 2 (described by scalar fields $\phi_1$ and $\phi_2$, respectively), as well as composites where a droplet enhanced in one species is encapsulated by another enhanced in the other species. The droplets are dynamic, both in terms of being randomly nucleated and dissolved, and their fluctuating shapes (see video S2 in the supplement). The effective non-reciprocal interaction reverses sign when the modulus of the order parameter is increased, i.e., while at low densities 1 chases 2, at higher densities 2 chases 1. The emergent imperfect PT symmetry breaking with local restoration of reciprocal interactions produces two new states, namely an effervescent wave which is a hybrid state with droplets and a traveling pattern, shown in Fig. \ref{fig:nrch_SS}(a-b), and effervescence without the traveling pattern, shown in Fig. \ref{fig:nrch_SS}(c-d).

\begin{figure}[t]
	\centering 
	\includegraphics[width= 0.95 \linewidth]{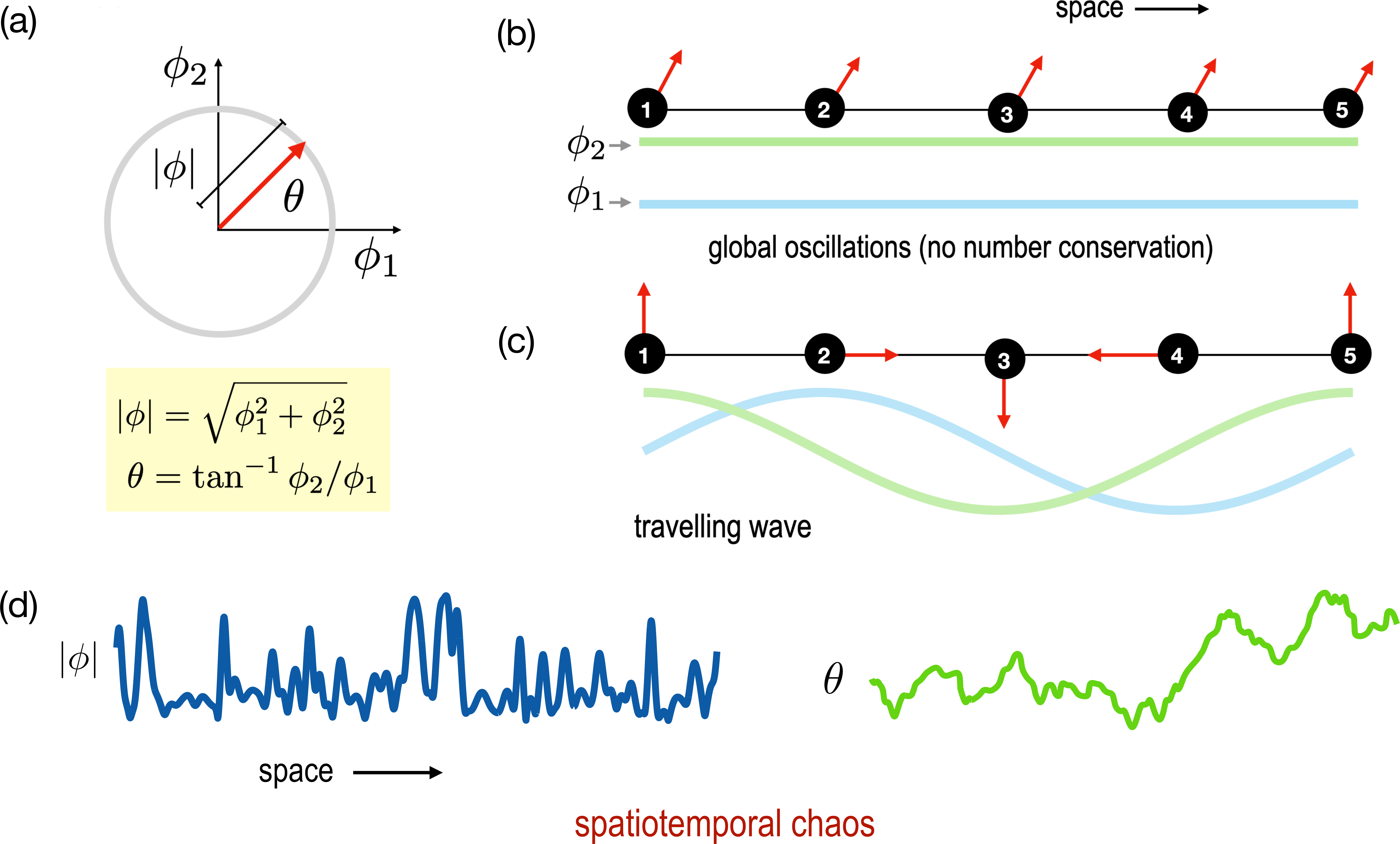}
	\caption{
	(a) The dynamics of two scalar fields $\phi_{1,2}$ in Eqs. \eqref{variantNRCH-1} and \eqref{variantNRCH-2} can be described alternatively in terms of the phase angle $\theta$, related to their relative concentration, and the modulus $|\phi|$, related to the net density. (b) In a system without number conservation, 
	a possible dynamical steady-state comprises global oscillations of uniform densities (green and blue line). 
	(c) Number conservation imposes a constraint that necessitates more a complicated synchronization pattern, which results in a travelling wave state with $\phi_{1,2}$ oscillating in space and time. 
	(d) Non-linear non-reciprocal interactions lead to spatiotemporal chaos in the steady-state where both $|\phi|$ and $\theta$ go through non-repetitive patterns; the snapshot corresponds to $\alpha_0 = \alpha_1 = 4$.}
	\label{fig:schematic}
\end{figure}

The phenomenon of effervescence reveals a granular structure for the domains that restore reciprocal interactions, as evident in the domain size distribution shown in Fig. \ref{fig:granule-size}(a-b). We observe a prominent peak for a fundamental {\em reciprocal granule} and an oscillatory pattern for larger areas [Fig. \ref{fig:granule-size}(a)] that originate from the composition of the domains being in the form of clusters of the reciprocal granules (see Supplemental Movies S1 and S2). For the effervescent waves, we observe a coexistence between the selected values for the modulus of the order parameter corresponding to the traveling pattern and the reciprocal granules, as observed in the distribution of the order parameter shown in Fig. \ref{fig:granule-size}(c). From the value of the modulus of the order parameter $|\phi|$ [see Fig. \ref{fig:schematic}(a)] one can verify that the effervescent granules correspond to a local restoration of PT symmetry. Effervescence gives rise to spatiotemporal chaos, and the emergence of an effective noise from the deterministic nonlinear dynamics, due to nonlinear non-reciprocal interactions. 


\begin{figure*}
	\centering
	\includegraphics[width= 0.95\linewidth]{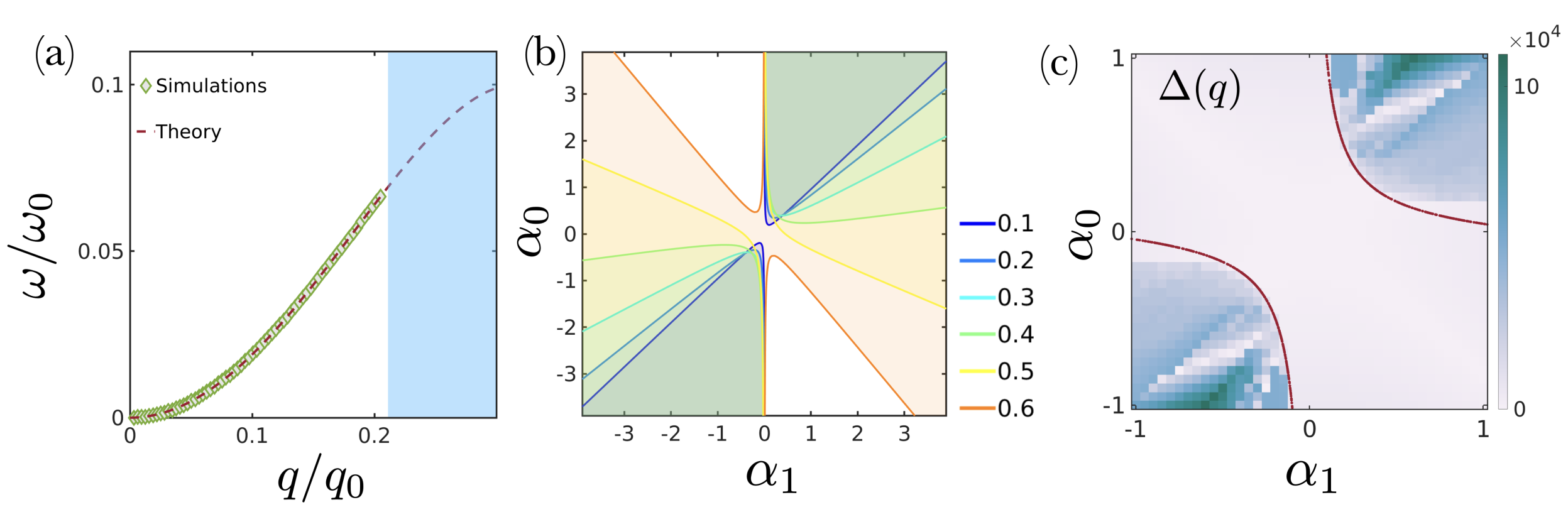}
	\caption{Dispersion and stability of travelling waves in the nonlinear NRCH model. (a) Comparison of the dispersion relation for $\alpha_0 = 1$ and $\alpha_1= -1$ as predicted by theory and as calculated from numerical simulations. The frequency is calculated from simulations with an initial condition $\phi^{\rm w}_{q}(\br,0)$ and a Fourier transform of the time series obtained at a randomly chosen fixed position in space. (b) The shaded area represents the unstable region of the phase space for travelling waves with different values of wavenumber $q/q_0$ (shown in legend) in the $(\alpha_0,\alpha_1)$ plane. (c) Theoretical predictions in panel (b) are checked using numerical simulations with initial conditions slightly perturbed $\phi^{\rm w}_{q}(\br,0)$ [see Eq. \eqref{eq:planeWave}]. Panel (c) shows a heat-map of $\Delta$ [see Eq. \eqref{eq:delta}]. The red line is for $q = 6 \pi/L$, where $L$ is the domain size (in units of $q_0^{-1}$).}
	\label{fig:stabilityAnalysis}
\end{figure*}

\section{Nonlinear non-reciprocal coupling} \label{sec:model}

To build our theoretical framework, we can start with the dynamics of conserved fields $\phi_{i}$ ($i=1,2$) that can be written as $\dot{\phi}_i = \Gamma_i \bm{\nabla}^2 \mu_i$, in terms of the scalar chemical potentials $\mu_i$ and mobilities $\Gamma_i$. At equilibrium, $\mu_i$ can be obtained from a free energy $F$ via $\mu_i \equiv \delta F / \delta \phi_i$. The free energy is chosen as $F = \int_{\bm{r}} f(\phi_i(\bm{r},t)) +\frac{K_1}{2} (\bm{\nabla} \phi_1)^2 + \frac{K_2}{2} (\bm{\nabla} \phi_2)^2$, 
where $f$ is the Helmholtz free energy (per unit volume) that describes phase separation in homogeneous systems. 
We now introduce non-equilibrium activity in the model by adding a non-reciprocal interaction between the two species. This can be achieved by introducing an anti-symmetric coupling between the species without any loss of generality, because the symmetric (reciprocal) part of the interaction can be absorbed in the expression for $f$. We can write the governing equations for the two fields as
\beq
\partial_t \phi_1 &=&  \Gamma_1 \bm{\nabla}^2 \left[  \frac{\partial f}{\partial \phi_1}  + \alpha(\phi_1,\phi_2) \phi_2 \right] -  \Gamma_1 K_1 \bm{\nabla}^4 \phi_1, \label{variantNRCH-1} \\
\partial_t \phi_2 &=&  \Gamma_2 \bm{\nabla}^2 \left[  \frac{\partial f}{\partial \phi_2} - \alpha(\phi_1,\phi_2) \phi_1 \right] -  \Gamma_2 K_2 \bm{\nabla}^4 \phi_2,
\label{variantNRCH-2}
\eeq
where the non-reciprocal coupling $\alpha(\phi_1,\phi_2)$ is taken to be a function of the fields, as a generalization of the NRCH model introduced in Ref. \cite{sahaPRX_2020}. We note the presence of number conservation is an important constraint that enhances the richness of the dynamics of such systems [see the examples in Figs. \ref{fig:schematic}(b-c)].

Equations \eqref{variantNRCH-1} and \eqref{variantNRCH-2} can be written equivalently as an equation for a complex field $\phi = \phi_1 + i \phi_2$ with an amplitude $|\phi| = \sqrt{\phi_1^2 + \phi_2^2}$ and a phase $ \theta = \tan^{-1} \phi_2/\phi_1$; see Fig. \ref{fig:schematic}(a). Dynamical steady-states resembling chemical turbulence are observed when $|\phi|$ and $\theta$ do not relax to stationary profiles [Fig. \ref{fig:schematic}(d)]. For the specific choice of the free energy density $f$, we consider the Landau-Ginzburg form \cite{HohenbergHalperinRMP}
\beq
f = -\frac{1}{2} |\phi|^2 + \frac{1}{4}|\phi|^4,
\label{eq:free_energy}
\eeq
which promotes phase separation into domains with $\phi_1,\phi_2 = \pm 1 $ at equilibrium, although we find that the steady states of \eqref{variantNRCH-1} and \eqref{variantNRCH-2} are independent of the detailed form of $f$, as expected. In the same spirit, we have chosen the following expression for $\alpha$ 
\beq
\alpha(\phi_1 , \phi_2)  =  \alpha_0 - \alpha_1 |{\phi}|^2,
\label{eq:alphaExp} 
\eeq
which introduces a non-linear nonreciprocal coupling between the two fields. For simplicity, we set $\Gamma_1 = \Gamma_2 = \Gamma$ and $K_1 = K_2  = K$. With the above choices, the nonlinear NRCH model is best described in terms of the dynamics of $\phi$, which satisfies the following equation
\beq
\partial_t \phi = \Gamma \bm{\nabla}^2 \left[- (1+ \iu \alpha_0) \phi + (1+ \iu \alpha_1) |\phi|^2 \phi  - K \bm{\nabla}^2 \phi\right], \nonumber \\ 
\label{eq:eqPhi-1}
\eeq
or equivalently, the dynamics of the amplitude and phase fields described earlier. We will now discuss the different dynamical steady-states of the non-linear NRCH model. 


\section{Travelling waves}\label{sec:wave}
We start by exploring the possibility of Eq. \eqref{eq:eqPhi-1} adopting travelling wave solutions, which is a natural consequence of the number conservation constraint as shown in Fig. \ref{fig:schematic}(c). 
Our choice of the bulk free-energy in Eq. \eqref{eq:free_energy} allows us to write down an exact dispersion relation for the travelling waves for a specific average composition of the system, which we choose as follows $\langle{\phi}_1 \rangle =\langle {\phi}_2 \rangle = 0$. The model is invariant under global phase rotation, i.e. the transformation $\phi \to e^{i\theta} \phi$ leaves \eqref{eq:eqPhi-1} unchanged. At this composition, the homogeneous state is linearly unstable to perturbations irrespective of the values $\alpha_{0,1}$. To capture the properties of the pattern formation process, we use a travelling wave trial solution $\phi^{\rm w}_{\bq}$ as parameterized by a wavenumber $\bq$, namely, 
\beq
\phi^{\rm w}_{\bq}(\br,t) = \rho_{\bq} \,e^{i( \bq \cdot \br - \omega t)},
\label{eq:planeWave}
\eeq
and substitute it in Eq. \eqref{eq:eqPhi-1} to obtain expressions for the amplitude $\rho_q$ and the dispersion relation $\omega(q)$. We find
\beq
\rho_{\bq} &=& \sqrt{1-\frac{q^2}{q_0^2}}, \;\; \forall \; q < q_0, \label{eq:rho-q} \\
\omega({\bq}) &=& {\Gamma} q^2 \left[- \alpha_0 + \alpha_1 \left( 1- \frac{q^2}{q_0^2} \right) \right], 
\label{eq:dispersion}
\eeq
where $q_0 = 1/\sqrt{K}$ [see Fig. \ref{fig:stabilityAnalysis}(a)]. 
The solutions in the form of Eq. \eqref{eq:planeWave} exist for all values of $\alpha_0$ and $\alpha_1$ and for a wide spectrum of wavelengths. It is, however, imperative to examine the stability of these solutions.

\begin{figure*}
	\centering
	\includegraphics[width= 1\linewidth]{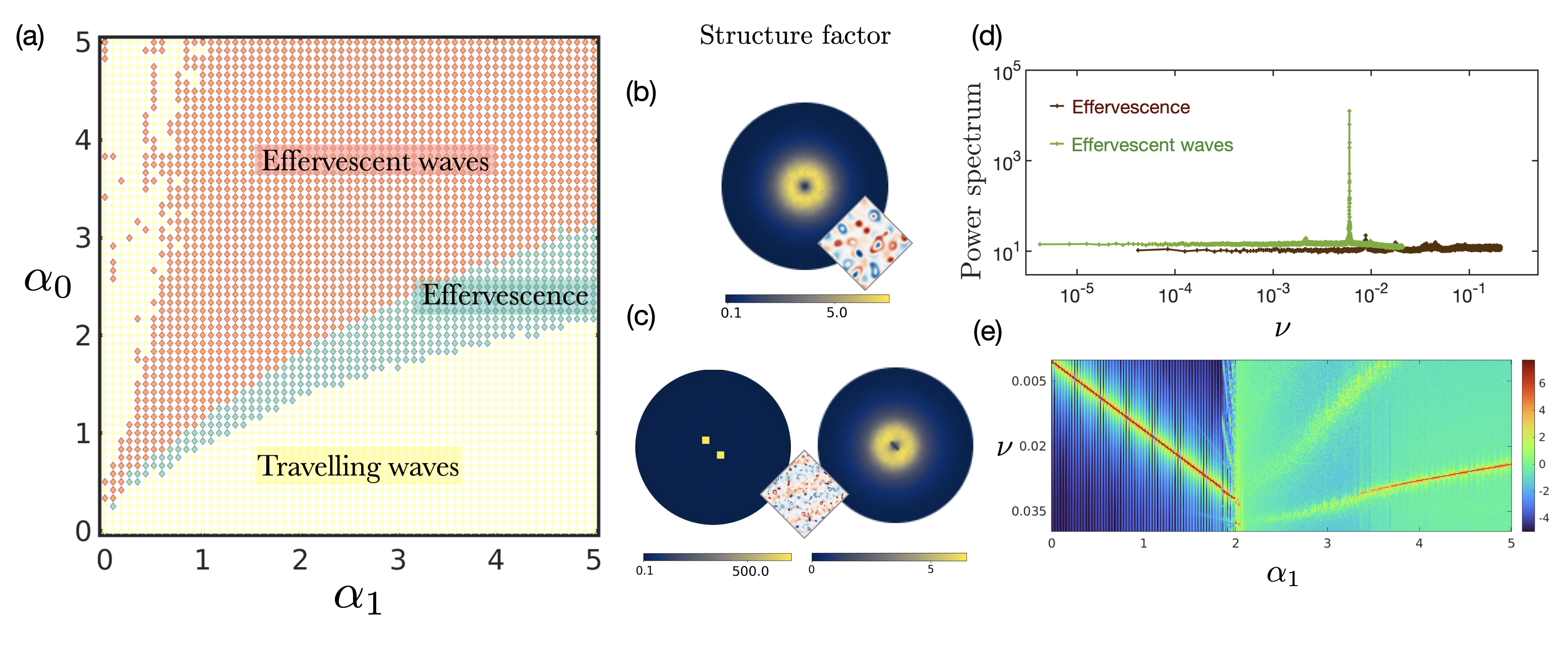}
	\caption{State diagram and correlation functions. (a) State diagram for the nonlinear NRCH model in two dimensions. Symbols indicate the points where the simulations were carried out and the colours correspond to the steady-state of the system. The steady states are summarized in Fig. \ref{fig:nrch_SS}. Panels (b-c) depict heat-maps for the structure factor $S(\bk)$ in the two states. (d) Power spectrum plotted as a function of the frequency $\omega$ for effervescent waves and effervescence for parameters $\alpha_0 = \alpha_1 = 4$, and $\alpha_0 =2$, $\alpha_1 = 4.33$, respectively. For effervescent waves, $S(\omega)$ shows a pronounced peak with a nearly constant background of temporal fluctuations. In the effervescent state, the spectrum is nearly constant and indistinguishable from white noise. In both cases, $S(\omega)$ decays at large frequencies. Panel (e) shows a heat map of the $S(\omega)$ as a function of $\alpha_1$ keeping $\alpha_0 = 2.5$. Note that the steep peak for travelling waves disappears in the effervescence case and reappears for effervescent waves. The dispersion is linear in both cases. The dotted lines are added to emphasise the sharp boundaries between the dynamical steady-states. }
	\label{fig:phaseDiagram}
\end{figure*}

To perform the stability analysis, we insert a trial solution of the form 
\begin{equation}
    \phi_{\bq}(\br,t) = [\rho_{\bq}+ \delta \rho_{\bq}(\br,t)] \, e^{i(\bq \cdot \br - \omega t)}, \label{eq:ansatz-lin-sta}
\end{equation}
in Eq. \eqref{eq:eqPhi-1}, and derive the effective governing equation for the perturbation $\delta \rho_{\bq}(\bk,t)$ in Fourier space with the wavenumber $\bk$, at the linear order, as has been done for the case of metachronal waves in cilia \cite{Meng2021}. The eigenvalues of the resulting linear dynamical equations in Fourier space can then be calculated (see Appendix \ref{appA:stability}) and used to isolate the dominant behaviour of the system as reflected in the eigenvalue with the larger real part. Using an expansion up to quadratic order in $\bk$, and a decomposition of the wavevector into the longitudinal component $k_L = \bk \cdot \bq/q$ and the transverse component $\bk_T = \bk \cdot ({\bm I}-\bq \bq/q^2)$, we can obtain the dominant eigenvalue, which we present as 
\begin{equation}
    \lambda(\bk) =  \iu V k_L  - D_L k_L^2 - D_T k_T^2 ,\label{eq:lambda-1}
\end{equation}
where the advection velocity $V$, and the longitudinal and transverse diffusion coefficients $D_L$ and $D_T$, are found as
\beq
V(q) &=& 2 \Gamma q \left(-\alpha_0 + \alpha_1   - 2 \alpha_1  {q^2}/{q_0^2} \right) ,\nonumber \\
D_L(q) &=& -{\Gamma} \alpha_1(\alpha_0 - \alpha_1)    + \frac{\Gamma \alpha_1^2 q^2 (3q^2 - 5q_0^2 )}{q_0^2(q_0^2-q^2)} \nonumber \\
 && + \frac{\Gamma q^2(q_0^2 - 3 q^2)}{q_0^2(q_0^2-q^2)} ,\nonumber \\
 D_T(q) &=& - \Gamma  \alpha_1(\alpha_0 - \alpha_1) + {\Gamma \alpha_1^2 q^2}/{q_0^2} .
\label{eq:stability}
\eeq
The travelling wave solutions in Eq. \eqref{eq:planeWave} are unstable in the part of the phase space where $D_L<0$. First, note that for $\alpha_1 =0$, $D_L$ reduces to the expression that holds for the conserved real Landau-Ginzburg dynamics, i.e. $\alpha_0$ alone does not create turbulence. For $q \to 0 $, $D_L \approx -\alpha_1(\alpha_0 - \alpha_1)$, indicating that an interplay of $\alpha_0$ and $\alpha_1$ is necessary for destabilizing plane waves. However, $\alpha_1$ alone can be used to tune $D_L$ to negative values at sufficiently large values of $q$. 

The stability diagram in the ($\alpha_0,\alpha_1$) plane is shown in Fig. \ref{fig:stabilityAnalysis}(b), with the unstable regions corresponding to $D_L<0$ being shaded (and the colours correspond to the wave-numbers indicated in the legend). For wavenumbers lower than a threshold value of $q_0/\sqrt{3}$, the unstable region consists of two unconnected pieces in the quadrants $\alpha_{0},\alpha_{1}>0$ and $\alpha_{0},\alpha_{1}<0$. Above the threshold, the two regions connect to form a single connected unstable region enclosing the origin. The Eckhaus stability criterion at equilibrium, which states that all wavelengths greater than $q_0/\sqrt{3} $ are unstable, thus determines the topology of the stability diagram.  

The result \eqref{eq:stability} is checked using numerical simulations with slightly perturbed travelling waves of a chosen wavelength as the initial condition and allowing the system to evolve for a sufficiently long time. The difference between the space averaged amplitude and the amplitude of the input wave defined as
\beq
\Delta(q) = \frac{1}{A} \int \mbox{d}^2 \br |{\phi}(\br,t)| - \rho_{\bq},
\label{eq:delta}
\eeq
is calculated in the ($\alpha_0,\alpha_1$) plane to determine the stability of the travelling waves; see Fig. \ref{fig:stabilityAnalysis}(c). Here $A=4 L^2$ is the area of the system. The wavelength of the sinusoidal wave, $q$, and the time periodicity, $\omega(q)$, of the wave at a fixed position in space, are determined using Fourier transforms. 

\section{State diagram} \label{sec:state-diag}
The state diagram presented in Fig. \ref{fig:phaseDiagram} is constructed in a single quadrant of the ($\alpha_0,\alpha_1$) plane, $\alpha_{0,1}>0$. From Eqs. \eqref{variantNRCH-1} and \eqref{variantNRCH-2}, it is clear that the simultaneous transformations $\alpha_{0} \to -\alpha_{0}$ and $\alpha_{1} \to -\alpha_{1}$ merely changes the sign of the fields, $\phi_{1,2} \to - \phi_{1,2}$. Therefore, it is sufficient to scan the dynamics in two adjacent quadrants only. However, we probe just the one for which both $\alpha_0, \alpha_1 > 0$ since for $\alpha_0<1$ and $\alpha_1>0$, the plane wave in Eq. \eqref{eq:planeWave} is a stable solution. 

We find that for $\sign(\alpha_0) = \sign(\alpha_1)$ and $|\alpha_1| \approx |\alpha_0|$, the steady-state is spatiotemporally chaotic, i.e. the scalar fields show non-repetitive oscillatory patterns. In these chaotic states, oscillating density fields coexist with granular domains formed by phase separation. We observe that the net non-reciprocal coupling, $\alpha(\phi_1,\phi_2)$, vanishes in parts of the space where $\phi_1^2+\phi_2^2 \approx 1$. Therefore, in domains where $|\phi| \approx 1$, reciprocity in interactions is locally restored, i.e. Eq. \eqref{eq:eqPhi-1} reduces to the conserved real Landau-Ginzburg dynamics, which results in the formation of non-motile phase-separated reciprocal granules.

The order parameter used to distinguish the travelling waves from the states with spatiotemporal chaos is the null density, which we define as the total number of simulation sites where $|\phi| \leq 0.1$. The null density is a good descriptor for determining the boundary between the travelling wave states [denoted with yellow markers in Fig. \ref{fig:phaseDiagram}(a)] and the effervescent states (denoted with green markers) as it jumps a few orders of magnitude upon crossing the boundary. Effervescence and effervescent waves [red markers in Fig. \ref{fig:phaseDiagram}(a)] are distinguished from one another by the structure factor, defined as
\beq
S(\bm{k})=\frac{1}{T} \int_0^T {\rm d} t \, {\phi}_i(\bm{k},t) {\phi}_i(-\bm{k},t),
\label{eq:structureFactor}
\eeq
where summation over repeated index $i$ is implied, and the power spectrum $S(\omega)$, defined as 
\beq
S(\omega) = \frac{1}{A} \int {\rm d}^2 \br \,  {\phi}_i(\bm{r},-\omega) {\phi}_i(\bm{r},\omega).
\label{eq:powerSpectrum}
\eeq
We observe that $S(\bk)$ is isotropic in the effervescent state [Fig. \ref{fig:phaseDiagram}(b)] and shows distinct peaks corresponding to the wavelength of the travelling wave in the effervescent-waves [Fig. \ref{fig:phaseDiagram}(c)]. Moreover, $S(\omega)$ exhibits a nearly constant plateau and is indistinguishable from white noise in the effervescent state, while a pronounced peak appears in addition to the nearly constant background for effervescent waves [Fig. \ref{fig:phaseDiagram}(d)]. We can also probe the heat-map of $S(\omega)$ as a function of $\omega$ and $\alpha_1$ for fixed $\alpha_0$ [Fig. \ref{fig:phaseDiagram}(e)]. We observe that the peak for travelling waves disappears in the effervescence case and reappears for effervescent waves. 
\begin{figure}
	\centering
	\includegraphics[width= 0.8\linewidth]{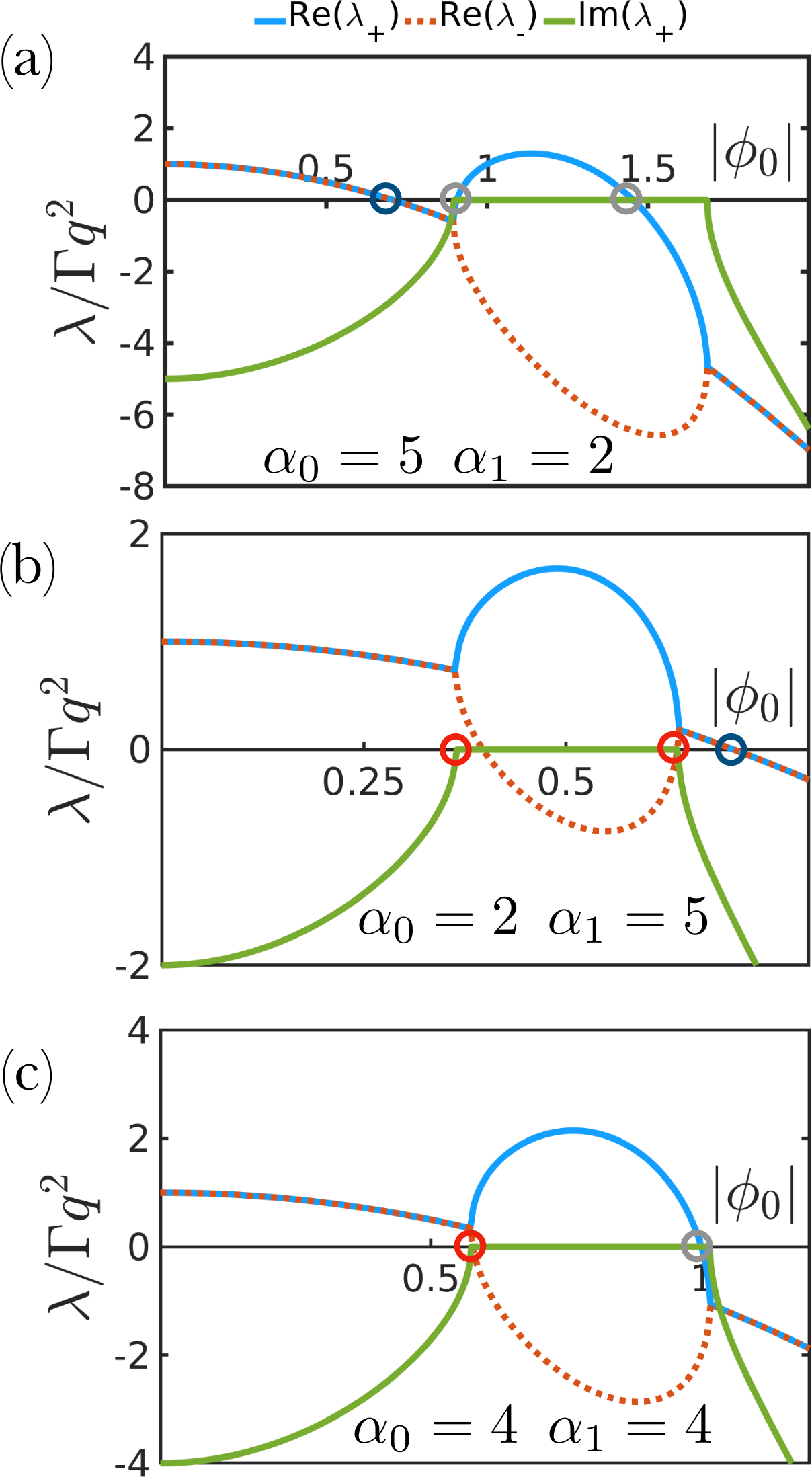}
	\caption{Eigenvalues as functions of the average composition $|\phi_0|$ for fixed values of $\alpha_0$ and $\alpha_1$. Panel (a-c) show the emergence of an exceptional point highlighted by the red circles. The Hopf bifurcation points are encircled in blue. }
	\label{fig:eigen}
\end{figure}

\section{Effect of composition}\label{sec:compos}

 \begin{figure*}[t]
	\centering
	\includegraphics[width= 1\linewidth]{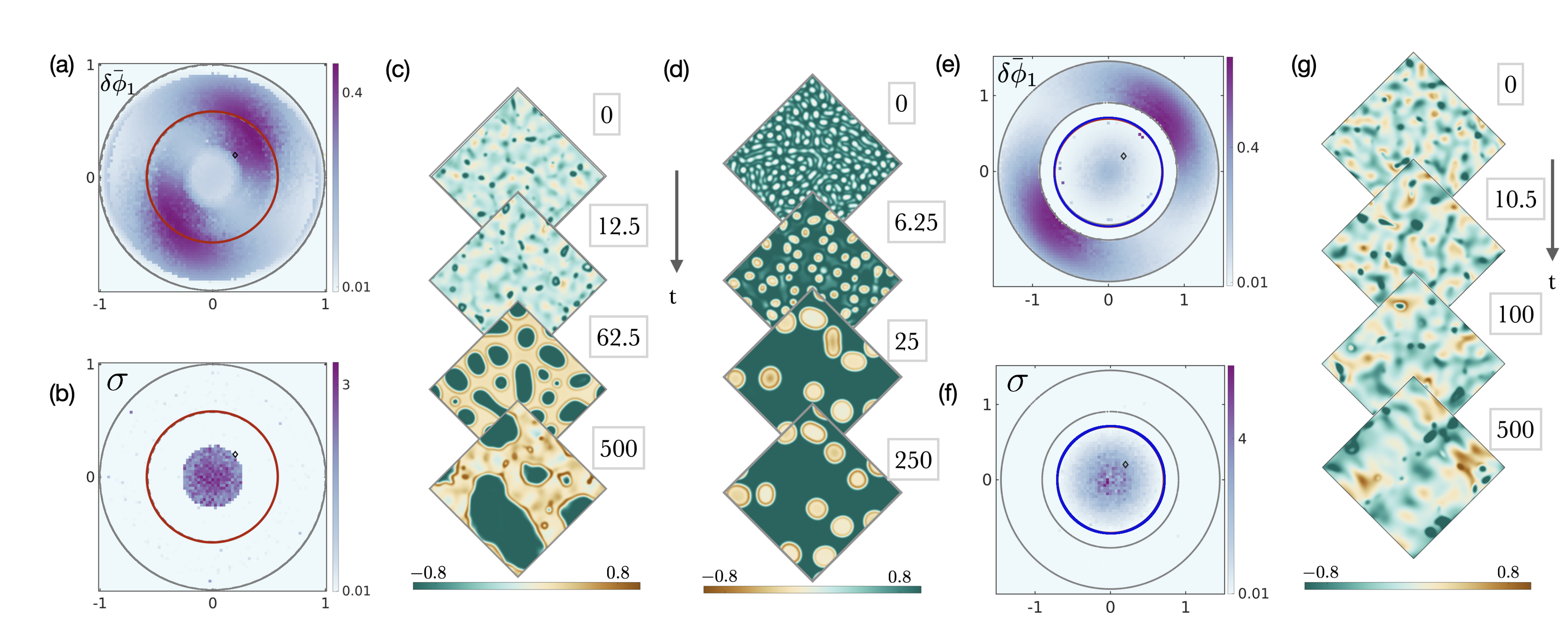}
	\caption{Effect of tuning the mean composition. (a-b) Linear stability diagram and its numerical verification in the composition plane ($\phi_1,\phi_2$) for $\alpha_0 = \alpha_1 = 4$. The solid red line denotes the exceptional points and the grey line marks the edge of the unstable region. Heat-maps of the quantities $\delta \bar{\phi}_1$ and $\sigma$ [see the discussion following \eqref{eq:exceptionalPoints}] calculated from simulations are superimposed on the stability diagrams in the panels (a) and (b) for comparison. The heat-map in (a) denotes the points where the homogeneous state is unstable while in (b) it shows $\sigma$, which probes the strength of the oscillatory states. Snapshots from simulations are shown in panels (c) and (d) for $\phi_0 = 0.25 + \iu 0.25$ and $\phi_0 = 0.4 + \iu 0$, respectively (see Supplemental Movies S3 and S4). The steady-state in both cases consists of phase separated droplets with oscillating boundaries. This is in contrast to the travelling bands with droplets observed for $\phi_0 = 0 + \iu 0$. (e-f) Numerical verification of the linear stability analysis for $\alpha_0 = 5$ and $\alpha_1 = 2$ using methods similar to panels (a-b). The blue line denotes the points where the Hopf bifurcation occurs. (g) Snapshots from simulations for $\phi_0 = 0.2 + \iu 0.2$ showing an effervescent wave.}
	\label{fig:Stability1}
\end{figure*}

We have so far kept the average composition fixed at $\langle \phi \rangle = 0$. We now discuss how the steady-states change as we tune the average composition. Note first that the invariance of the free energy under unitary rotations in the composition plane enabled us to determine the exact solution \eqref{eq:planeWave} for $\langle \phi \rangle = 0$. This invariance is broken when we move away from $\langle \phi \rangle = 0$ in the composition plane. In order to check the stability of the homogeneous state with average composition $\phi_0 \equiv \langle \phi_1 \rangle + \iu \langle \phi_2 \rangle$, we perturb the expression in Eq. \eqref{eq:eqPhi-1} around $\phi_0$ and obtain the following linearized equations of motion for small deviations $\delta \phi$ using the dynamical matrix $\mathcal{D}$
\beq
\begin{bmatrix} \delta \dot{\phi} \\ \delta \dot{\phi^*} \end{bmatrix} = 
\begin{bmatrix} \mathcal{D}_{11}  & \mathcal{D}_{12} \\ \mathcal{D}_{21} & \mathcal{D}_{22} \end{bmatrix}  
\begin{bmatrix}
\delta \phi \\
\delta \phi^*
\end{bmatrix},
\label{eq:matrixeq-1}
\eeq
where
\beq
\mathcal{D}_{11} &=& \mathcal{D}_{22}^* = \Gamma q^2 \left [(1+i\alpha_0) - 2(1+i\alpha_1) |{\phi}_0|^2 \right], \nonumber \\
\mathcal{D}_{12} &=& \mathcal{D}_{21}^* = - \Gamma q^2 (1+i\alpha_1)  {\phi}_0^2,
\eeq
up to order $q^2$. The eigenvalues $\lambda_{\pm}$ of the non-Hermitian stability matrix $\mathcal{D}$ in Eq. \eqref{eq:matrixeq-1} can be complex. For the stability of the homogeneous state, real parts of $\lambda_{\pm} = \mbox{tr}(\mathcal{D})/2 \mp \frac12 \sqrt{\mbox{tr}(\mathcal{D})^2 - 4 \mbox{det}(\mathcal{D}) }$ should be negative. Complex values for $\lambda_{\pm}$ imply that a slightly perturbed mixed state develops an oscillatory instability and the system generally evolves into a steady-state that carries signatures of these oscillations like the travelling wave or those summarized in Fig. \ref{fig:nrch_SS}. The determinant and the trace of the $\mathcal{D}$, given as 
\beq
&& \mbox{det}(\mathcal{D}) =\Gamma^2 q^4 \left [ (1-2|\phi_0|^2)^2 + (\alpha_0 - 2\alpha_1 |\phi_0|^2)^2 \right. \nonumber \\
&& \left.-|\phi_0|^4 (1+\alpha_1^2)\right], \nonumber \\
&& \mbox{tr}(\mathcal{D}) = \Gamma q^2 \left [2 - 4|\phi_0|^2 \right],
\label{eq:DetTrace}
\eeq
are functions of $|\phi_0|$ only. Since $\lambda_{\pm}$ depend on the trace and determinant of $\mathcal{D}$, the stability is thus determined by $|\phi_0|$ alone. For complex values of $\lambda_{\pm}$, the real and imaginary parts are
\beq
\mbox{Re}(\lambda) &=& \Gamma q^2 \left [ 1 - 2|\phi_0|^2 \right], \nonumber \\
\mbox{Im}(\lambda) &=& \Gamma q^2 \left [(\alpha_0 - 2 \alpha_1 |\phi_0|^2)^2 - |\phi_0|^4 (1+\alpha_1^2) \right].
\label{eq:eigen}
\eeq
For $|\phi_0| = 0$, we observe that $\mbox{Re}(\lambda)>0$ and $\mbox{Im}(\lambda) \neq 0$ independently of $\alpha_{0,1}$. Therefore, in the middle of the unstable region the homogeneous state develops oscillatory instabilities in response to small perturbations. The real and imaginary parts of the eigenvalues are plotted as functions of $|\phi_0|$. Two types of bifurcation points arise as $|\phi_0|$ is changed while keeping other parameters constant (as shown in Fig. \ref{fig:eigen}): Hopf bifurcation where the real parts of a pair of complex eigenvalues change sign, and exceptional point where the two $\lambda_{\pm}$ coalesce while the corresponding eigenvectors are parallel. Upon crossing an exceptional point, the eigenvalues develop imaginary parts. The nonlinearity of the non-reciprocal parameter being considered enhances the richness of the stability diagram. For $|\phi_0| < 1/\sqrt{2}$, two exceptional points appear at the following values of $|\phi_0|$:
\beq
|\phi_0|^2 = \frac{2 \alpha_0 \alpha_1}{3 \alpha_1^2-1} \pm \frac{\alpha_0 \sqrt{\alpha_1^2+1}}{3 \alpha_1^2-1}.
\label{eq:exceptionalPoints}
\eeq
A third possibility occurs for $|\phi_0| = 1/\sqrt{2}$ and $\alpha_0 = \alpha_1 \pm \frac12 \sqrt{1+\alpha_1^2}$ where $\mbox{Re}(\lambda) = \mbox{Im}(\lambda) = 0$. Finally, a pair of real eigenvalues could both change sign signalling an instability where perturbations grow and lead to the formation of a bulk separated state.

The results of the stability analysis are verified numerically by running 81 $\times$ 81 simulations keeping $\alpha_{0,1}$ fixed and by varying $\phi_{1,2}$ (see Fig. \ref{fig:Stability1}). For $\alpha_0 = 5$ and $\alpha_1=2$, $\phi_{1,2}$ are varied in the range $-1$ to $1$ while for $\alpha_0 = 4$ and $\alpha_1=4$ they are varied in the range $-1.5$ to $1.5$. The average deviation from the homogeneous state $\bar{\delta \phi_1} \equiv \frac{1}{A}\int {\rm d}^2\br \left[\phi_1-\mbox{Re}(\phi_0)\right]$ is calculated in the steady state to identify the points where the homogeneous system is unstable. To identify the oscillatory steady state we calculate the area enclosed in ($\phi_1,\phi_2$) space by the boundary enclosing the trajectory $(\phi_1(\br,t),\phi_2(\br,t))$ at a constant $\br$, namely, $\sigma(\br)=\frac12 \oint \phi_1 {\rm d}\phi_2- \phi_1 {\rm d}\phi_2$. We observe that $\bar{\delta \phi_1}>0$ where $\mbox{Re}(\lambda_{\pm}) > 0 $, while we obtain nonzero values of $\sigma$ where $\mbox{Im}(\lambda_{\pm}) > 0$, indicating oscillatory steady-states. 

Moving away radially from the centre $|\phi_0| = 0$ in the composition space ($\phi_1,\phi_2$), the currents driving the phase separation appear to dominate over the non-reciprocal interactions. For $\alpha_0 = \alpha_1=4$, the effervescent waves change into a predominantly phase separated state with domains spanning the system size and with fluctuating interfaces (see Supplemental Movies S3 and S4). For $\alpha_0 = 5$ and $\alpha_1=2$, the effervescent waves persist until the very edge of the region beyond which the homogeneous state is stable. 



\section{Concluding Remarks}\label{sec:conc}
We have introduced a model with nonlinear non-reciprocal interactions between two species, and studied the phase separation dynamics of the system and its dynamical steady states. We have observed a new type of chemical spatiotemporal chaos that arises due to imperfect breaking of PT symmetry, involving fluctuating domains in space where the symmetry is temporarily restored. This effect produces the startling phenomena of effervescence and effervescent traveling waves. We also observe that our model exhibits fluctuations that can act as a background effective white noise due to the non-reciprocal nonlinearities, similarly to the case of Kuramoto-Sivashinsky equation. 

The non-reciprocal coupling $\alpha$ and the free energy $f$ were chosen such that we can obtain an analytical form for the travelling waves first reported in \cite{sahaPRX_2020,You_Marchetti_2020}, thereby establishing the stability of the waves. The emergence of spatiotemporal chaos is attributed to the non-reciprocal coupling that changes sign as a function of the amplitude of scalar fields. A linear stability analysis enables us to highlight the interplay between the two non-reciprocal coefficients, which destabilizes the traveling waves. we have verified that the results presented here hold quite generally, independently of the choice of the bulk free energy $f$, and also for two alternative forms of $\alpha$. 

The effect of non-reciprocal interactions in the presence of number conservation constraint, which is characteristic of many active matter systems, leads to the emergence of novel dynamical states. We have also highlighted the role of composition in tuning the pattern-forming behaviour of the system, which enhances the connection of the model to bulk phase separating systems.

Our work sheds light onto the rich and complex behaviour that can arise in minimal models of active matter system with non-reciprocal interactions. We hope that our work will pave the way for new studies of the role of non-reciprocity in colloidal systems with tunable interactions \cite{LavergneBechinger2019} or in the field of swarm robotics \cite{Scholz2018}.

\begin{acknowledgments}
We acknowledge fruitful discussions with Jaime Agudo-Canalejo. This work has received support from the Max Planck School Matter to Life and the MaxSynBio Consortium, which are jointly funded by the Federal Ministry of Education and Research (BMBF) of Germany, and the Max Planck Society.
\end{acknowledgments}

\appendix

\section{Linear Stability Analysis}\label{appA:stability}

To investigate the stability of the plane wave solution \eqref{eq:planeWave} to the nonlinear NRCH equation \eqref{eq:eqPhi-1}, we insert the form \eqref{eq:ansatz-lin-sta} into Eq. \eqref{eq:eqPhi-1}, and expand the equation up to the first order in $\delta \rho_{\bq}$ in Fourier space, taking into account the definitions given in Eq. \eqref{eq:dispersion}. We obtain
\beq 
\begin{bmatrix}
	\delta \dot{\rho}_{\bq}(\bk) \\\\
    \delta \dot{\rho}^*_{\bq}(\bk)
\end{bmatrix}
= \begin{bmatrix} \mathcal{M}_{11} & \mathcal{M}_{12} \\\\ \mathcal{M}_{21} & \mathcal{M}_{22} \end{bmatrix}  
\begin{bmatrix}
    \delta {\rho}_{\bq}(\bk) \\\\
    \delta {\rho}^*_{\bq}(\bk)
\end{bmatrix}, 
\label{eq:matrixeq}
\eeq
where 
\beq
&& \mathcal{M}_{11}(\bq,\bk) = i {\Gamma} q^2 \left[- \alpha_0 + \alpha_1 \left( 1- \frac{q^2}{q_0^2} \right) \right] - \Gamma K (\bk-\bq)^4 \nonumber \\
&& \hskip1cm + \Gamma \left[(1+i \alpha_0) - 2 (1+i \alpha_1) \left( 1- \frac{q^2}{q_0^2} \right) \right] (\bk-\bq)^2, \nonumber \\ \nonumber \\
&& \mathcal{M}_{12}(\bq,\bk) = -\Gamma (1+i \alpha_1)\left( 1- \frac{q^2}{q_0^2} \right) (\bk-\bq)^2 , \nonumber \\ \nonumber\\
&& \mathcal{M}_{21}(\bq,\bk) = -\Gamma (1-i \alpha_1)\left( 1- \frac{q^2}{q_0^2} \right)(\bk+\bq)^2 , \nonumber \\ \nonumber\\
&& \mathcal{M}_{22}(\bq,\bk) = -i {\Gamma} q^2 \left[- \alpha_0 + \alpha_1 \left( 1- \frac{q^2}{q_0^2} \right) \right]  - \Gamma K (\bk+\bq)^4 \nonumber \\
&& \hskip1cm + \Gamma \left[(1-i \alpha_0) - 2 (1-i \alpha_1)\left( 1- \frac{q^2}{q_0^2} \right) \right] (\bk+\bq)^2.\nonumber
\eeq
The stability of the plane wave solution \eqref{eq:planeWave} is determined by the eigenvalues of the $\mathcal{M}$, which are given as $\lambda_{\pm} \equiv \mbox{tr} \mathcal{M} /2 \mp \frac12 \sqrt{\mbox{tr} \mathcal{M} ^2 - 4 \mbox{det} \mathcal{M}}$. To the zeroth order in $k$, the eigenvalues are $\lambda_{-} = -2 q^2 \left( 1-{q^2}/{q_0^2} \right)$ and $\lambda_{+} = 0$.  For small wavelength perturbations $k \ll q$, the branch of eigenvalues $\lambda_{-}$ remains negative, and thus stabilizing, and the stability of the travelling waves is determined by $\lambda_{+}$ alone. We calculate $\lambda_{+}$ up to quadratic order in $k^2$ to probe the advection and diffusion effects. To $O(k^2)$, we obtain the results presented in Eqs. \eqref{eq:lambda-1} and \eqref{eq:stability}.

\bibliography{biblio}

\end{document}